\documentclass[useAMS,usenatbib,fleqn]{mn2e}
\usepackage{graphicx,times,subfigure}
\usepackage{bm}
\usepackage{epsf}
\usepackage{amsmath}
\usepackage{amssymb}
\usepackage{cases}
\usepackage{float}
\usepackage[caption = false]{subfig}
\usepackage{colortbl}
\usepackage{multirow}

\def\apjs{Astrophysical Journal, Supplement}

\def\aap{Astronomy and Astrophysics}
\def\apjl{Astrophysical Journal, Letters}

\def\la{\langle}

\def\be{\begin{equation}}
\def\ee{\end{equation}}
\def\ben{\begin{eqnarray}}
\def\een{\end{eqnarray}}

\def\mathhMpc{h \mbox{Mpc}^{-1}}

\begin{document}
\title[Testing Modified Gravity with Cosmic Shear]{Testing Modified  Gravity  with Cosmic Shear}
\author[Harnois-D\'eraps et al.]
{J. Harnois-D\'eraps$^{1}$, D. Munshi$^{2}$, P. Valageas$^{3,4}$, L. van Waerbeke$^1$, P. Brax$^{3,4}$,  P. Coles$^{2}$, \newauthor L. Rizzo$^{3,4}$\\
$^{1}$Department of Physics and Astronomy, University of British Columbia, 6224 Agricultural Road, Vancouver, B.C., V6T 1Z1, Canada\\
$^{2}$ Astronomy Centre, School of Mathematical and Physical Sciences, University of Sussex, Brighton BN1 9QH, United Kingdom\\
$^{3}$ CEA, IPhT, F-91191, Gif-sur-Yevette, Ce\'dex, France\\
$^{4}$ CNRS, URA, 2306, F-91191, Gif-sur-Yevette, Ce\'dex, France\\ }

\maketitle
\begin{abstract}
We use the cosmic shear data from the Canada-France-Hawaii Telescope Lensing Survey to
place  constraints on $f(R)$ and {\it Generalized Dilaton} models of modified gravity.
This is highly complimentary to other probes since the constraints mainly come from the non-linear scales:
maximal deviations with respects to the General-Relativity + $\Lambda$CDM scenario occurs at $k\sim1 h \mbox{Mpc}^{-1}$. 
At these scales, it becomes necessary to account for known degeneracies with baryon feedback and massive neutrinos,
hence we place constraints jointly on these three physical effects.
To achieve this, we formulate these modified gravity theories within a common tomographic parameterization,
we compute their impact on the clustering properties relative to a GR universe,
and propagate the observed modifications into the weak lensing $\xi_{\pm}$ quantity.
Confronted against the cosmic shear data,  we reject the $f(R)$ $\{ |f_{R_0}|=10^{-4}, n=1\}$ model with more than 99.9\% confidence interval (CI) when assuming a  $\Lambda$CDM dark matter only model.
In the presence of baryonic feedback  processes and massive neutrinos with total mass up to 0.2eV, the model is disfavoured with {\it at least} 94\% CI  in all different combinations studied.  
Constraints on the $\{ |f_{R_0}|=10^{-4}, n=2\}$ model are weaker, but nevertheless disfavoured with at least 89\% CI.
We identify several specific combinations of neutrino mass, baryon feedback and $f(R)$ or Dilaton gravity models that are excluded by the current cosmic shear data.
Notably,  universes with three massless neutrinos and no baryon feedback are strongly disfavoured in all modified gravity scenarios studied.
These results indicate that competitive constraints may be achieved with future cosmic shear data.

\end{abstract}
\begin{keywords}: Cosmology-- Modified Gravity Theories -- Methods: analytical, statistical, numerical
\end{keywords}

\section{Introduction}
Explaining the late-time acceleration of the Universe first reported in \citet{Ries98, Perl99} represents a major challenge in modern cosmology,
and current interpretations mostly rely on the inclusion of dark energy components and/or modifications to the theory of General Relativity (GR).
One important difficulty encountered in solving this puzzle relates to the fact that,
by construction, the background dynamics in viable dark energy and modified gravity models are almost indistinguishable \citep{Bertschinger:2006aw,SHS07,Brax:2008hh}.
These two frameworks only really decouple when considering the evolution of matter density fluctuations and of perturbations associated with the metric.
In addition, there are various ways in which a modification of gravity on large scales could account for the apparent acceleration \citep{Clift12,JJKT14}.
Exploiting this, many observational probes based on large scale structure formation have been proposed to test theories of modified gravity,
including galaxy clustering \citep{PS08,OLH}, integrated Sachs-Wolfe effect in the cosmic microwave background (CMB) anisotropies
and its cross-correlation with galaxy density \citep{SPH07}, cluster abundance \citep{Jain:2007yk,Lombriser:2010mp},
peculiar velocities \citep{Li:2012by, 2014MNRAS.444.3926, 1504.06885}, redshift-space distortions from spectroscopic surveys \citep{Guzz08,Jennings:2012pt, Asaba13},
21cm observations \citep{Hall13} and weak gravitation lensing \citep{HKV07,Sch08,TT08, Simpson13, Wilcox15}.

In this paper we investigate the extent to which current weak lensing surveys can constrain departures from GR.
In particular, we study the signatures of two specific classes of parametrized modified gravity theories, the $f(R)$ and the generalized Dilaton models,
on the cosmic shear measurement extracted from the CFHTLenS  \citep{2013MNRAS.433.2545E}.
These models are known to cause an enhancement of structure formation over scales in the range [0.2 - 20] Mpc$h^{-1}$, an effect which could be detectable with current  lensing surveys.
In addition, the departure of these models from General Relativity is maximal at scales of $k\sim1 h^{-1}\mbox{Mpc}$, 
which are difficult to interpret with other clustering data due to the  large uncertainty in the galaxy bias.
This makes the weak lensing approach special, probing modified gravity models at the scale of influence of the `fifth' force.

In its approach, this paper is an extension of  \citet[hereafter HWVH]{2014arXiv1407.4301H},
where the same data was used to place joint constraints on baryon feedback models and on the sum of neutrino mass.
The general idea can be understood as follows: on the one hand, the accuracy achieved by modern CMB experiments \citep{2013ApJS..208...19H, 2013arXiv1303.5076P} on most $\Lambda$CDM parameters is
at the percent level; on the other hand, the modified gravity effects we are looking for affect  the baseline signal by  up to 20 percent at small scales.
 It is therefore justified to assume a fixed cosmology and search for possible deviations.
 Any residual uncertainty in the cosmology can thereafter be treated as systematic uncertainty in the analysis.
While next generation weak lensing experiments such as RCSLenS\footnote{http://www.rcslens.org}, DES\footnote{http://www.darkenergysurvey.org},
KiDS\footnote{http://kids.strw.leidenuniv.nl},  Euclid\footnote{ http://sci.esa.int/euclid} and LSST\footnote{http://www.lsst.org/lsst} will have enough statistical power to repeat this analysis in a full MCMC pipeline, we demonstrate here that we can find interesting results
with simpler tools and existing data.

This paper is organized as follows: In \textsection\ref{sec:theory} we review the theoretical formulation of
structure formation in $f(R)$ and Dilaton gravity theories;
\textsection\ref{sec:lensing} describes the theoretical and numerical modelling of the weak lensing signal,
and details our cosmic shear measurement from the CFHTLenS data. In \textsection\ref{sec:results} we present and discuss  our results,  and conclude in \textsection\ref{sec:conclu}.
The baseline cosmological parameters that are used throughout our study correspond to the {\it WMAP}9 + BAO + SN $\Lambda$CDM cosmology:
$h=0.6898$, $\Omega_{\rm m}=0.2905$, $\Omega_{\Lambda}=0.7095$, $\Omega_{K}=0$, $w=-1$, $\sigma_8=0.831$ and $n_s=0.969$.
The reason why we did not opt for the {\it Planck} cosmology is to minimize the effect of the known cosmological tension in our model rejection strategy.
Otherwise this would involve a full MCMC calculation including all cosmological parameters and both data sets as in \citet{MacCrann}, which is not necessary in our approach.  
In the end however,  we do marginalize over this cosmological discrepancy.


\section{Modified Gravity Theories}
\label{sec:theory}

Modified theories of gravity can be distinguished by their screening properties in dense environments. 
Indeed, given the strong Solar System constraints, these theories need to have a built-in screening mechanism,
suppressing the deviations from GR. Three types of such mechanisms have emerged in the last few years: the {\it Chameleon}, {\it K-mouflage} and {\it Vainshtein} models \citep[see][for a comparison  between these different screening mechanisms]{Brax:2014wla}.
On the one hand, K-mouflage and Vainshtein models involve non-linear kinetic terms describing additional  scalar fields whose presence modifies GR predictions. 
On the other hand, modifications of the Chameleon type can be broadly categorized as either containing additional couplings between the metric and new scalar fields,
or involving extra geometric terms. These two equivalent descriptions can be captured by the {\it tomographic} parameterization, which will be used throughout this paper \citep{BDL12,Brax:2012gr}. 


In all Chameleon cases, modifications of gravity induce a global enhancement of the effective force of gravity, due to the `fifth force',
which directly translates into an increase of structure formation.
In this Section, we review two different types of modified gravity, namely the {\it Dilaton} 
and the $f(R)$ models; we describe their distinct screening mechanisms, and detail 
their parameterization in the context of large scale structure formation.

\subsection{Gravity in Dilaton models}
\label{sec:Dilatons}

The {\it Dilaton} and {\it Symmetron}\footnote{We do not further investigate the Symmetron, K-mouflage nor Vainshtein models in this paper.} theories of modified gravity are Chameleon models that exhibit
the {\it Damour-Polyakov} property \citep{DP94}, according to which the coupling between the scalar field $\varphi$ and the rest of the matter components approaches zero in
dense environments \citep{Piet05, OP08, HK10}.
In contrast to the case of $f(R)$ theories, described in Sec.~\ref{sec:f(R)} below,
the scalar field here takes on a small mass everywhere and thus mediates a long-range (screened) force.

These Dilaton models are scalar-tensor theories, where the action defining the system
takes the general form
\ben
S & = & \int d^4x \sqrt{-g} \left[ \frac{M_{\rm Pl}^2}{2} R - \frac{1}{2} (\nabla\varphi)^2 
- V(\varphi) - \Lambda_0^4 \right] \nonumber \\
&& + \int d^4x \sqrt{-\tilde{g}} \tilde{\cal L}_m( \psi^{(i)}_m,\tilde{g}_{\mu\nu}) ,
\label{eq:S-dilaton-def}
\een
where $M_{\rm Pl}=(8\pi {\cal G}_{\rm N})^{-1/2}$ is the reduced Planck mass (in natural units),
$\Lambda_0^4$ is the cosmological constant term  today,
$g$ is the determinant of the Einstein-frame metric tensor $g_{\mu\nu}$ and
$\tilde{g}$ the determinant of the Jordan-frame metric tensor $\tilde{g}_{\mu\nu}$.
The two metrics are connected via a conformal rescaling
\ben
\tilde{g}_{\mu\nu} = A^2(\varphi) g_{\mu\nu} .
\label{conformal}
\een

The various matter fields $\psi^{(i)}_m$ are governed by the Jordan-frame
Lagrangian density $\tilde{\cal L}_m$ and the scalar field $\varphi$ by the
Einstein-frame Lagrangian density ${\cal L}_{\varphi}=-1/2(\nabla\varphi)^2-V(\varphi)$, where $V(\varphi)$ is the potential of the scalar field\footnote{In equation (\ref{eq:S-dilaton-def}), 
we explicitly added the cosmological
constant term $\Lambda_0^4$, so that the minimum of $V(\varphi)$ is zero and
is reached for $\varphi \rightarrow \infty$. Alternatively, this term could also be interpreted
as the non-zero minimum of the scalar field potential.}.
There is no explicit coupling between matter and the scalar fields, and the fifth
force on matter particles due to $\varphi$ arises from the conformal
transformation given by equation (\ref{conformal}) (more precisely, through gradients of $A$).

In the original Dilaton model, the potential $V(\varphi)$ and the 
coupling\footnote{This coupling is often defined as $A(\varphi) = 1 + {1 \over 2} {A_2 \over M^2_{\rm Pl}} (\varphi - \varphi_*)^2$, 
where $ \varphi_*$ is some free parameter of the model. We opted  to absorb $\varphi *$ into $\varphi$ in equation (\ref{A-phi-def}), 
a choice that has no physical impact anyway.}  
$A(\varphi)$ with the metric have the following functional forms:
\ben
 V(\varphi) = V_* \exp \left( - {\varphi \over M_{\rm Pl}} \right) ,
\label{V-exp-dilaton} 
\een
\ben
 A(\varphi) = 1 + {1 \over 2} {A_2 \over M^2_{\rm Pl}} \varphi^2 ,
\label{A-phi-def}
\een
where $\{V_*, A_2 \}$ are the two free parameters.
In regions where  $\varphi \approx 0$, the coupling to matter is negligible,
and gravity converges to GR.
However,  the field nevertheless mediates a long range gravitational force that has an effect  elsewhere, i.e. in less dense environments.
This model can be generalized to a greater class of Dilaton models, by keeping the coupling
function as in equation (\ref{A-phi-def}) but considering more general potentials.
Then, instead of specifying the model by its potential $V(\varphi)$, it is 
re-casted in the tomographic parametrization $\{ \beta(a), m(a) \}$ in terms of the cosmological scale factor $a(t)$,
 where the coupling $\beta(a)$ and the scalar field mass $m(a)$ are defined as \citep{BDL12, BV13}:
\ben
\beta(a) \equiv \beta[\bar{\varphi}(a)] = M_{\rm Pl} \frac{d\ln A}{d\varphi}(\bar{\varphi}), 
\label{eq:beta-a-def} 
\een
\ben
m^2(a) \equiv m^2[\bar{\varphi}(a),\bar{\rho}(a)] = \frac{1}{c^2} \left[
\frac{d^2V}{d\varphi^2}(\bar{\varphi}) + \bar{\rho} \frac{d^2 A}{d\varphi^2}(\bar{\varphi}) 
\right].
\label{eq:m-a-def}
\een
Hereafter, we denote with an overbar unperturbed cosmological background quantities,
and with a subscript `0' quantities evaluated today. 
For instance, $\bar{\rho}(a)= {3 \Omega_{\rm m0} H_0^2 M_{\rm Pl}^2}/{a^3}$ is the background matter density,
$\bar{\varphi}$ is the mean value of the field, $H_0$ is the current value of the Hubbles parameter, and $\Omega_{\rm m0}$ is the current matter density.
Also, $c$ is the speed of light in vacuum.
In this paper we consider the simple forms
\ben
m(a)=m_0 \, a^{-r} , \;\;\; \beta(a)=\beta_0 \exp\left[-s \frac{a^{2r-3}-1}{3-2r} \right],
\label{eq:symm}
\een
with
\ben
s = \frac{9 A_2 \Omega_{\rm m0}H_0^2}{c^2 m_0^2 }.
\label{eq:s-def}
\een
In this framework,  the Yukawa potential given by equation (\ref{V-exp-dilaton}) corresponds to $r=3/2$.
The values of the free parameters $\{m_0, r, \beta_0, s\}$ that enter equation (\ref{eq:symm})  
are displayed in Table \ref{tabular:tab2}.
The models \{A, B, C, D\} were chosen such as to correspond to those studied in 
\citet{BV13} and \citet{BDLWZ12}, where detailed comparisons between numerical 
and analytical calculations are presented.
More specifically, the models \{A, B, C\} probe the dependence on $\{s,\beta_0,r\}$ respectively,
all other parameters being fixed, while models D probe the dependence on $m_0$
at fixed $A_2$. 
We added the models E that probe the dependence on the parameter $m_0$
at fixed $\{s,\beta_0,r\}$.
These models probe deviations from the $\Lambda$CDM cosmology of less than $20\%$, in terms
of the matter power spectrum.
Let us now explore the detailed mechanism through which the power spectrum of matter fluctuations
is affected by this theory.

In these Dilaton models, the coupling function $A$ is always very close to unity,
so that most Einstein-frame and Jordan-frame quantities (e.g., Hubble expansion rates
or densities) are almost identical.
Indeed, using $|\bar{A}-1| \ll 1$, we can see from equations (\ref{A-phi-def}) and
(\ref{eq:beta-a-def}) that  $\bar{A} \simeq 1 + \beta^2/(2 A_2)$.
From equation (\ref{eq:s-def}) we also obtain $A_2 \sim (c m_0/H_0)^2$.
Solar System tests of gravity such as that analysed in \citet{Chiba03} imply that $m_0 \gtrsim 10^3 H_0/c$, whence
$A_2 \gtrsim 10^6$ and
\ben
| \bar{A} - 1 | \lesssim 10^{-6} .
\label{eq:bar-A-bound}
\een
Therefore, the Jordan-frame and Einstein-frame scale factors and background matter 
densities, related by $\tilde{a} = \bar{A} a$ and $\tilde{\bar\rho} = \bar{A}^{-4} \bar{\rho}$,
can be considered equal, as well as the cosmic times and Hubble expansion rates. 
In the rest of this Section we work in the Einstein frame, where the analysis of the gravitational
dynamics are simpler.

In the Einstein frame, the Friedmann equation takes the usual form
\ben
3 M_{\rm Pl}^2 H^2 = \bar{\rho} + \bar{\rho}_{\varphi} + \bar{\rho}_{\Lambda} ,
\label{eq:Friedmann-dilaton}
\een
where we explicitly separate contributions from the matter ($\bar{\rho}$) and scalar field ($ \bar{\rho}_{\varphi}$) components and from  the cosmological
constant $\bar{\rho}_{\Lambda}$.
The background value of the scalar field potential is given by
\ben
\frac{d \bar{V}}{d\bar{\varphi}} + \frac{\beta}{M_{\rm Pl}} \bar{\rho} = 0 .
\label{eq:background-KG}
\een
Combining with equation (\ref{eq:m-a-def}), and writing $m=m(a)$, this leads to
\ben
\frac{d\bar{\varphi}}{d a} = \frac{3 \beta \bar{\rho}}{c^2 M_{\rm Pl} a m^2} , \;\;\;
\frac{d\bar{V}}{d a} = - \frac{3\beta^2\bar{\rho}^2}{c^2 M_{\rm Pl}^2 a m^2} ,
\een
whence
\ben
\frac{\dot{\bar\varphi}^2}{2\bar{\rho}} \sim \left( \frac{H}{c m} \right)^4 \sim 10^{-12} ,
\;\;\; \frac{\bar{V}}{\bar{\rho}} \sim  \left( \frac{H}{c m} \right)^2 \sim 10^{-6} .
\een
Thus, the scalar field energy density is dominated by its potential term, which is negligible
as compared with the matter density.
Therefore, the Friedmann equation (\ref{eq:Friedmann-dilaton}) is governed by the matter
density and the cosmological constant and we recover the $\Lambda$CDM cosmological
expansion, $3 M_{\rm Pl}^2 H^2 = \bar{\rho} + \bar{\rho}_{\Lambda}$,
up to an accuracy of $10^{-6}$.

We now briefly consider the behaviour of metric and density fluctuations.
In the quasi-static limit, the scalar field is given by the Klein-Gordon equation,
\ben
\frac{c^2}{a^2} \nabla^2 \varphi = \frac{d V}{d\varphi} + \rho \frac{d A}{d\varphi} ,
\een
and at linear order over the matter density and scalar field fluctuations we obtain
\ben
\frac{\delta\varphi}{M_{\rm Pl}} = - \frac{\beta}{c^2 M_{\rm Pl}^2} 
\frac{\delta\rho}{m^2+k^2/a^2} ,
\label{eq:dphi-scaling}
\een
where $k$ is the comoving wave number. Using equation (\ref{eq:background-KG})
this gives
\ben
| \delta A | \sim \frac{|\delta \rho|}{\bar{\rho}} \left( \frac{H}{c m} \right)^2 
\frac{1}{1+k^2/a^2m^2} \lesssim 10^{-6} ,
\label{eq:dA-scaling}
\een
so that the perturbations of the conformal factor $A^2$ are negligible compared
to unity. Also, 
\ben
\frac{\delta \rho_{\varphi}}{\delta\rho} \sim 
\left( \frac{H}{c m} \right)^2 \frac{1}{1+k^2/a^2m^2} \lesssim 10^{-6} ,
\label{eq:drho-phi-scaling}
\een
hence fluctuations of the scalar field energy density are negligible 
compared with the matter density fluctuations.

Therefore, the main source that drives modifications to structure growth is not a different background evolution,
nor perturbations in the scalar field energy density, but really the action of the fifth force on the matter field.
In the Newtonian gauge, the perturbed metric can be written as
\begin{eqnarray}
ds^2 = - (1 + 2\Phi) dt^2 + a^2(t) (1-2\Psi) \delta_{ij}dx^idx^j ,
\label{eq:perturbed-metric}
\end{eqnarray}
where $\Phi$ and $\Psi$ are the Einstein-frame metric gravitational potentials.
Using equations (\ref{eq:dphi-scaling}) and (\ref{eq:drho-phi-scaling}), we can check that the
impact of the scalar field fluctuations on the metric potentials are negligible, and 
we have within a $10^{-6}$ accuracy
\ben
\Phi = \Psi = \Psi_{\rm N} ,
\label{eq:Phi-Psi-PsiN-dilaton}
\een
where $\Psi_{\rm N}$ is the Newtonian potential given by the Poisson equation,
\begin{eqnarray}
\frac{\nabla^2}{a^2} \Psi_{\rm N} = 4\pi {\cal G}_{\rm N} \delta\rho
= \frac{3 \Omega_{\rm m0} H_0^2}{2 a^3} \delta.
\label{eq:Poisson-GR}
\end{eqnarray}
In the above expression,  $\delta=\delta\rho/\bar{\rho}$ is the matter density contrast.
However, the dynamics of matter particles are modified by the scalar field, which 
gives rise to the fifth force given by ${\bf F} = - c^2 \nabla \ln A$.
 That is, in the Euler equation  we must add a
fifth-force potential, $\Psi_A=c^2 \ln A$, that is not negligible.
When solving for structure growth given the parameters listed in Table~\ref{tabular:tab2}, the new term can lead to $10-20\%$ deviations in the matter density power spectrum.

\subsection{{Gravity in $f(R)$ theories}}
\label{sec:f(R)}

In models based on $f(R)$ gravity, the Einstein-Hilbert action 
is modified by promoting the Ricci scalar $R$ to a function of $R$  \citep{Buch70,Staro80,Staro07,HS07}.
The new action $S$ in $f(R)$ gravity theories can be written as:
\begin{eqnarray}
S = \int d^4 x \sqrt{-g} \, \left[ \frac{M^2_{\rm Pl}}{2} [ R + f(R) ] - \Lambda_0^4  
+ {\cal L}_m(\psi_m^{(i)}) \right] ,
\label{S-fR-def}
\end{eqnarray}
where we explicitly added the cosmological constant contribution\footnote{The terms $R$ and $\Lambda_0^4$
 are often included within the function $f(R)$. Written in the form of equation (\ref{S-fR-def}), $f(R)$ describes
 deviations from both GR and the $\Lambda$CDM cosmology.}.
The $f(R)$ models are most easily described in the Jordan frame, which is why,
in this Section, we denote with a tilde Einstein-frame quantities instead of Jordan-frame ones,
contrary to the notation of \textsection\ref{sec:Dilatons}.
In the parameterization of \citet{HS07}, the functional form $f({R})$ can be expressed in the high curvature limit as:
\begin{eqnarray}
f({R}) =  - {f_{R_0} \over n}{{R}_0^{n+1} \over {R}^n}, 
\;\;\; f_R \equiv {df(R) \over dR} = f_{R_0} \frac{R_0^{n+1}}{R^{n+1}} .
\label{eq:fofR}
\end{eqnarray}
The two independent parameters, $f_{R_0}$ and $n$, can be constrained by observations.
In the above expression, $R_0$ is the present value of the  Ricci scalar for the cosmological background.
Note that this parametrization and that of \citet{Staro07} both reproduce the same results in the large curvature regime.

The $f(R)$ theories of gravity also invoke the { Chameleon} mechanism to screen modifications of GR in dense environments such as in our Solar System.
Specifically, this occurs by requiring that all extra terms vanish in high curvature environment, such that  $f(|R| \gg |R_0|) \rightarrow 0$.
The background expansion otherwise follows the $\Lambda$CDM dynamics
and the growth of structure is only affected on intermediate and quasi-linear scales.

There is an essential connection between the formulation of the $f(R)$ theory presented above 
and scalar-tensor  theories of modified gravity.
Upon the coordinate rescaling $\tilde  g_{\mu\nu}= A^{-2}(\varphi)g_{\mu\nu}$
(recall that in this Section $\tilde{g}_{\mu\nu}$ is the Einstein-frame metric) 
with $A(\varphi)= \exp[\beta \varphi/ M_{\rm Pl}]$ and $\beta = {1/ \sqrt 6}$,
the $f(R)$ modifications to GR are re-casted as arising from contributions of an extra scalar field $\varphi$,  subject to a potential $V(\varphi)$ given by:
\begin{eqnarray}
V(\varphi) = \frac{M^2_{\rm Pl}}{2} \left( \frac{R f_R -f(R)}{(1+f_R)^2} \right) ,
\;\;\; f_R= \exp \left [ -\frac{2\beta\varphi}{M_{\rm Pl}} \right ] - 1 .
\end{eqnarray}
In that sense,  $f(R)$ theories are equivalent to a scalar-tensor theory expressed in the  Einstein frame \citep{Chiba03,NS04}.
In this new formulation,  the screening mechanism takes another form:
the mass of the scalar field grows with matter density, and a Yukawa-like potential suppresses the fifth force in dense environments. This can be conveniently reformulated by saying that
screening takes place wherever the scalar field  is small compared to the ambient Newtonian potential.

It turns out that all Chameleon-like models such as $f(R)$ theories can again be parameterized by the value of the mass $m(a)$ and the coupling $\beta (a)$ of the scalar field, in terms 
of the scale factor $a$ and the associated background matter density 
$\bar{\rho}(a)$.
With the specific functional form of $f(R)$ given by equation (\ref{eq:fofR}), we can  directly 
relate  $\{n , f_{R_0} \}$ to $\{\beta(a), m(a) \}$ via:
\begin{eqnarray}
m(a)=m_0  \left( \frac{4\Omega_{\Lambda 0} +\Omega_{\rm m0} a^{-3}}
{4\Omega_{\Lambda 0} +\Omega_{\rm m0}} \right)^{(n+2)/2}, \nonumber \\ 
m_0 = \frac{H_0}{c} \sqrt{ \Omega_{\rm m0} + 4 \Omega_{\Lambda 0}\over (n+1) | f_{R_0} |  } , \;\;\; \beta(a) = \frac{1}{\sqrt{6}} . 
\label{eq:m0_fr}
\end{eqnarray}
In this paper, we consider values of  $n=\{1,2\}$ and $|f_{R_0}| = \{10^{-4},10^{-5},10^{-6}\}$.
The larger value of $|f_{R_0}|$ is currently  ruled out by other independent  probes, so this serves as  a consistency test.
The numerical values for $\{\beta(a), m(a)\}$ corresponding to these three
models are listed in Table \ref{tabular:tab2}.

As for the Dilaton models described in \textsection\ref{sec:Dilatons}, the $f(R)$ models
that we consider in this paper follow very closely the $\Lambda$CDM cosmology
at the background level, mainly because $|f_{R_0}| \ll 1$.
Indeed, from the action (equation \ref{S-fR-def}) one obtains the Friedmann equation as \citep{Tsuji07};
\ben
3 M_{\rm Pl}^2 \left[ H^2 - \bar{f}_R (H^2+\dot{H}) + \bar{f}/6 
+ \bar{f}_{RR} H \dot{\bar{R}} \right] =  \bar{\rho} + \bar{\rho}_{\Lambda} ,
\label{Friedmann-fR}
\een
where the dot denotes the derivative with respect to cosmic time $t$ and
$f_{RR} =d^2f/dR^2$. In the background we have $\bar{R} = 12 H^2+6 \dot{H}$
and we can check that all extra terms in the brackets in equation (\ref{Friedmann-fR}) 
are of order $|f_{R_0}| H^2$, so that we recover the $\Lambda$CDM expansion,
$3 M_{\rm Pl}^2 H^2 = \bar{\rho} + \bar{\rho}_{\Lambda}$,
up to an accuracy of $10^{-4}$ for $|f_{R_0}| \lesssim 10^{-4}$.
Moreover, the conformal factor $A(\varphi)$ is given by $A=(1+f_R)^{-1/2}$,
so that $|\bar{A}-1| \lesssim 10^{-4}$ and the background quantities associated with
the Einstein and Jordan frames can be considered equal (and equal to the
$\Lambda$CDM reference) up to an accuracy of $10^{-4}$.

Considering the metric and density perturbations, we can again write the 
Newtonian gauge metric as in equation (\ref{eq:perturbed-metric}) (but this is now the
Jordan-frame metric). Then, in the small-scale sub-horizon limit
$k/a \gg H/c$, the modified Einstein equations lead to \citep{TT08}
\ben
&& \frac{\nabla^2}{a^2} \Phi = - \frac{c^2\nabla^2}{2a^2} \delta f_R 
+ 4\pi{\cal G}_{\rm N} \delta\rho , \label{Phi-fR-def} \\
&& \frac{\nabla^2}{a^2} \Psi = \frac{c^2\nabla^2}{2a^2} \delta f_R 
+ 4\pi{\cal G}_{\rm N} \delta\rho , \label{Psi-fR-def}
\een
where $\delta f_R= f_R - \bar{f}_F$ and $\delta\rho=\rho-\bar{\rho}$.
Therefore, in terms of the Newtonian gravitational potential $\Psi_{\rm N}$ defined
as in GR by equation (\ref{eq:Poisson-GR}), we have
\ben
\Phi = \Psi_{\rm N} - \frac{c^2}{2} \delta f_R , \;\;\;
\Psi = \Psi_{\rm N} + \frac{c^2}{2} \delta f_R .
\label{Phi-Psi-fR}
\een
Thus, because we work in the Jordan frame (in contrast with the Dilaton case
presented in \textsection\ref{sec:Dilatons}), the modification of gravity directly appears
through the metric potentials.
The fluctuations of the new degree of freedom $\delta f_R$
are given by:
\ben
\frac{3c^2\nabla^2}{a^2} \delta f_R = \delta R - 8\pi{\cal G}_{\rm N} \delta\rho .
\een
Finally, the dynamics of the matter particles is given by the geodesic equation,
where the Newtonian potential that appears in GR is replaced by the potential $\Phi$ given in
equation (\ref{Phi-Psi-fR}).

\begin{table*}
\begin{center}
\caption{Parameters describing the modified gravity theories considered in our study, 
mapped on the $\{\beta(a), m(a)\}$ surface, parameterized with $\{m_0, r, \beta_0, s\}$ 
following equation (\ref{eq:symm}).
The first five rows correspond to different realizations of the generalized Dilaton theories.
The last two rows show  $f(R)$ theories with $n=1$ and $2$ respectively, in which $m_0$ is given by equation (\ref{eq:m0_fr}), while $r\equiv 3(n+2)/2$, $\beta_0=1/\sqrt{6}$ 
and $s=0$. }
\begin{tabular}{| c | c | c | c | c | }
\hline
\hline
Model & $m_0 (h / \rm{Mpc})$ & $r$ & $\beta_0$ & $s$ \\
\hline
\hline
(A1, A2, A3) & $(0.334,0.334,0.334)$ & $(1.0,1.0,1.0)$ & $(0.5,0.5,0.5)$ & $(0.6,0.24,0.12)$ \\
\hline
(B1, B3, B4) & $(0.334,0.334,0.334)$ & $(1.0,1.0,1.0)$ & $(0.25,0.75,1.0)$ & $(0.24,0.24,0.24)$ \\
\hline
 (C1, C3, C4) & $(0.334,0.334,0.334)$ & $(1.33,0.67,0.4)$ & $(0.5,0.5,0.5)$ & $(0.24,0.24,0.24)$ \\
\hline
(D1, D3, D4) & $(0.667,0.167,0.111)$ & $(1.0,1.0,1.0)$ & $(0.5,0.5,0.5)$ & $(0.06,0.96,2.16)$ \\
\hline
(E1, E3, E4) & $(0.667,0.167,0.111)$ & $(1.0,1.0,1.0)$ & $(0.5,0.5,0.5)$ & $(0.24,0.24,0.24)$ \\
\hline
\hline
$n=1, \log_{10} |f_{R_0}| $ = (-4, -5, -6) & $(0.042,0.132,0.417)$ & $(4.5,4.5,4.5)$ & $(0.408,0.408,0.408)$ & $(0,0,0)$ \\
\hline
$n=2, \log_{10} |f_{R_0}| $ = (-4, -5, -6) & $(0.034,0.108,0.340)$ & $(6.0,6.0,6.0)$ & $(0.408,0.408,0.408)$ & $(0,0,0)$ \\
\hline
\hline
\end{tabular}
\label{tabular:tab2}
\end{center}
\end{table*}

\section{Weak Lensing}
\label{sec:lensing}

\subsection{Theory}
\label{sec:lensing-theory}

\subsubsection{Weak lensing convergence power spectrum}
\label{sec:convergence-power}

In all the cosmologies considered in this paper, we work in the Newtonian
gauge with the perturbed metric given by equation (\ref{eq:perturbed-metric}),
where $\Phi$ and $\Psi$ are the metric gravitational potentials\footnote{In the Dilaton models, this is understood as the Einstein-frame metric while in the
$f(R)$ models this is the Jordan-frame metric, following the approach described
in \textsection\ref{sec:theory}. In any case, we can work in either frame as the observational results
do not depend on this computational choice.}.
In practice, we measure the statistical properties of weak lensing distortions
by summing over many galaxy images.
This means that the measured signal is an integral over selected sources with a broad redshift distribution
$n(z_s) dz_s$ (mapped to $n(\chi) d\chi$ in terms of the radial distance, given the Jacobian $d\chi/dz$) 
that we normalize to unity.
Thus, introducing the kernel $g(\chi)$ that defines the radial depth of the survey:
\begin{eqnarray}
g(\chi) = \int_{\chi}^{\infty} d\chi_s n(\chi_s) \frac{\chi_s - \chi}{\chi_s} ,
\label{g-chi-def}
\end{eqnarray}
the integrated convergence field at a position ${\boldsymbol \theta}$ on the sky reads as:
\begin{eqnarray}
\kappa({\boldsymbol \theta}) = \int_0^{\infty} d\chi \frac{\chi}{c^2} g(\chi) 
\nabla^2 \Phi_{\rm wl}(\chi,\chi{\boldsymbol \theta}) .
\label{eq:kappa}
\end{eqnarray}
We assumed a flat background universe in the above equation, and introduced the weak lensing potential, defined by
\begin{eqnarray}
\Phi_{\rm wl} = \frac{\Phi + \Psi}{2} ,
\label{eq:Phi-wl-def}
\end{eqnarray}
which is a convenient when computing weak lensing modifications to GR. 
Solving equation (\ref{eq:kappa}) in multipole space and taking the ensemble average
of the squared complex norm, we obtain the convergence power spectrum: 
\begin{eqnarray}
C^{\kappa}_{\ell} = \int_0^{\infty} d\chi \frac{g(\chi)^2}{c^4} \frac{\ell^4}{\chi^4} 
P_{\Phi_{\rm wl}}(\ell/\chi;z)
\label{eq:Cl-Pkphi}
\end{eqnarray}
as an integral over the weak lensing power spectrum $P_{\Phi_{\rm wl}}(k;z)$. 
Note that the above also assumes both Limber and Born approximations.
From this, we also derive predictions for the cosmic shear two-point correlation functions 
$\xi_{\pm}(\theta)$, computed as:
\begin{eqnarray}
\xi_{\pm}(\theta) = \frac{1}{2\pi} \int C_{\ell}^{\kappa} J_{0/4}(\ell\theta)  \ \ell \ d\ell
\label{eq:xi}
\end{eqnarray}
where $J_{0/4}(x)$ are Bessel functions of the first kind.

\subsubsection{$C^{\kappa}_{\ell}$ in General Relativity}
\label{sec:Ckappa-GR}

In the $\Lambda$CDM cosmology + GR case, we can exactly express the weak lensing convergence
power spectrum (\ref{eq:Cl-Pkphi}) in terms of the total matter 
power spectrum $P(k)$ via Poisson equation.
Indeed, we can safely neglect the anisotropic stress,
and General Relativity gives:
\begin{eqnarray}
\Phi_{\rm wl} = \Phi = \Psi = \Psi_{\rm N} ,
\label{Phi-wl-Phi-Psi-PsiN-GR}
\end{eqnarray}
where $\Psi_{\rm N}$ is the Newtonian potential given by Poisson equation
(equation \ref{eq:Poisson-GR}).
Therefore, we recover
\begin{eqnarray}
P_{\Phi_{\rm wl}}(k;z) = \left( \frac{3 \Omega_{\rm m0} H_0^2}{2 a k^2} \right)^2 P(k;z) ,
\label{eq:Pphi-Pdelta}
\end{eqnarray}
and the convergence power spectrum (\ref{eq:Cl-Pkphi}) becomes:
\begin{eqnarray}
C^{\kappa}_{\ell} = \int_0^{\infty} d\chi W(\chi)^2 P(\ell/\chi;z) ,
\label{eq:limber}
\end{eqnarray}
with
\begin{eqnarray}
W(\chi) = \frac{3 \Omega_{\rm m0} H_{0}^{2}}{2 c^2} g(\chi) (1 + z) .
\label{eq:W-def}
\end{eqnarray}

\subsubsection{$C^{\kappa}_{\ell}$ in theories of modified gravity}
\label{sec:Ckappa-dilaton}


For the Dilaton models, we have seen in equation (\ref{eq:Phi-Psi-PsiN-dilaton}) and in \textsection\ref{sec:Dilatons} that
the two Einstein-frame metric potentials are equal to the Newtonian potential up to order $10^{-6}$ accuracy,
and that  background cosmological quantities such as
the Hubble expansion rate and the radial comoving distances are equal to those of the
$\Lambda$CDM reference within that same accuracy.
This means that equations (\ref{Phi-wl-Phi-Psi-PsiN-GR})-(\ref{eq:Pphi-Pdelta}) apply as in GR,
and that $C^{\kappa}_{\ell}$   is again given by 
equations (\ref{eq:limber})-(\ref{eq:W-def}).
Therefore, in terms of this weak lensing statistics, the modification of gravity
and the departures from the $\Lambda$CDM+GR results only appear through the
modified matter density power spectra $P(k;z)$, which 
we describe in \textsection\ref{subsubsec:MG_bias}.





In the case of  $f(R)$ models, we have seen in equation (\ref{Phi-Psi-fR}) that the two Jordan-frame
potentials are different from the Newtonian potentials, receiving contributions from terms linear in $\delta f_R$. 
However, these two extra terms exactly cancel in the weak lensing potential
(equation \ref{eq:Phi-wl-def}) such that $\Phi_{\rm wl} = \Psi_{\rm N}$.
Therefore, we recover equations (\ref{eq:Pphi-Pdelta})-(\ref{eq:W-def}) in $f(R)$ models too.
Moreover, we have seen that both the Jordan-frame and Einstein-frame background
quantities are equal to the reference $\Lambda$CDM background quantities
up to an accuracy of $10^{-4}$ for $|f_{R_0}| \lesssim 10^{-4}$.
This means that  weak lensing statistics
can again be computed in the reference background cosmology, 
so long as the modified matter density power spectrum is used.

\subsection{Non-linear matter power spectrum}

The choice of non-linear power spectrum to insert in equation (\ref{eq:limber}) depends on the cosmology under investigation.
In this paper, we are interested in constraining modified gravity models, but with respect to a $\Lambda$CDM baseline,
these are strongly degenerate with universes that include baryon feedbacks and/or massive neutrinos.
In the context of cosmic shear, these phenomenas are therefore intrinsically connected and must be jointly analysed.
We detail in this Section how we combine all these effects in the construction of our theoretical predictions.

\subsubsection{Dark matter only}

The first choice we make concerns the dark matter model $P^{\rm DM}(k)$, which is a delicate issue that has been thoroughly investigated in HWVH in a very similar context.
Following this work, we choose the dark matter only model that best reproduces the results from a number of $N$-body simulations,
then implement the combined effect of modified gravity, baryon feedback and massive neutrinos  relative to this dark matter only baseline.
Our dark matter only prediction is a hybrid model that combines the Extended Cosmic Emulator \citep{2013arXiv1304.7849H} with the recalibrated {\small HALOFIT} code by \citet{2012ApJ...761..152T}.
Its convergence properties have been well examined in HWVH, and it was shown to have the best agreement with independent high-resolution simulation suites,
compared with other models. In addition, HWVH examined the scatter across multiple models, and
estimated the theoretical uncertainty on the global dark matter only prediction for $\xi_{\pm}$. In this paper, we also incorporate this
model uncertainty in the analysis pipeline, at the level of the $\chi^2$ calculation (see \textsection\ref{subsec:error}).

\subsubsection{Neutrino and baryon feedback}

Following HWVH, we model the impact of massive neutrinos and baryon feedback on the matter power spectrum
as separate effects that can be expressed with multiplicative feedback terms, namely:
\begin{eqnarray}
  P^{{\rm DM}+\nu+b(\rm m)}(k,z) = P^{\rm DM}(k,z) \times b^2_{M_{\nu}}(k,z) \times b^2_{\rm m}(k,z). 
\end{eqnarray}
The underlying assumption is that both biases are independent, which is reasonable since baryons were found to have a one percent effect on the neutrinos bias
for $k<8 \mathhMpc$ \citep{2012MNRAS.420.2551B}.

We compute the {\it neutrino feedback bias} term $b^2_{M_{\nu}}$ with the {\small CAMB } cosmological code \citep{Lewis:1999bs},
which is reported to be accurate to better than 10 percent at $k = 10 \mathhMpc$ \citep{2012MNRAS.420.2551B}.
We assume one massive neutrino flavour, and fix the cosmology at high redshift -- i.e. we keep the primordial amplitude $A_s$ fixed  but let $\sigma_8$ vary.
We justify this choice from the fact that the former quantity is measured very accurately by CMB observations,
whereas our estimation of the latter quantity is much less accurate due to galactic and cluster bias.
We  construct the neutrino bias as
\begin{eqnarray}
b^2_{M_{\nu}}(k,z) \equiv \frac{P^{{\rm DM}+M_{\nu}}_{\rm CAMB}(k,z)}{P^{\rm DM}_{\rm CAMB}(k,z)},
\label{eq:nu_feedback}
\end{eqnarray}
where the $M_{\nu}$ (= $0.0$, $0.2$, $0.4$, or $0.6$ eV) superscript specifies the total neutrino mass considered,
and the subscript `{\small CAMB}' specifies that both quantities are measured from this cosmological numerical code.

The {\it baryonic feedback bias} is estimated
from two hydrodynamical simulations ran in the context of the OverWhelmingly Large (OWL) Simulation Project \citep{2010MNRAS.402.1536S}.
The dark matter only run (DM-ONLY) is a purely collision-less $N$-body calculation and acts as the baseline for this baryon feedback measurement only.
The AGN simulation run contains gas dynamics with physical prescriptions for cooling, heating, star formation and evolution, chemical enrichment, supernovae feedback and
active galactic nuclei feedback
\citep[see][for details about these simulations]{2011MNRAS.415.3649V}.
Following \citet{2011MNRAS.415.3649V, 2011MNRAS.417.2020S}, we measure the baryonic feedback bias by taking the ratio between the AGN and the DM-ONLY models\footnote{
The power spectrum measurements from the OWL simulation suite are publicly available at: {\tt http://vd11.strw.leidenuniv.nl}}:
\begin{eqnarray}
 b^2_{\rm m}(k,z) \equiv \frac{P^{{\rm DM}+b(\rm m)}_{\rm OWL}(k,z)}{P^{\rm DM}_{\rm OWL}(k,z)},
 \label{eq:b_feedback}
\end{eqnarray}
where the index $b(m)$ refers to either DM-ONLY or AGN, and the subscript `OWL' specifies that these quantities were measured specifically from the OWL simulation suite.

\begin{figure}
\centering
  \includegraphics[width=2.2in]{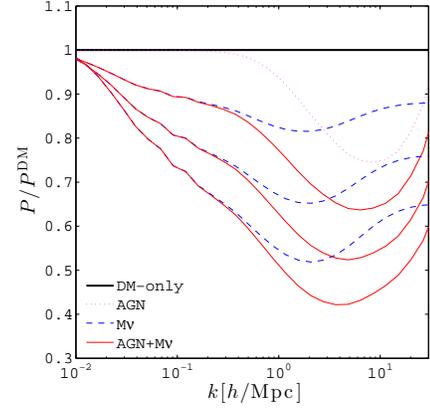} 
\vspace{-0.25cm}
\caption{ Combined effect from baryon feedback and massive neutrinos on the  matter power spectrum $P(k)$, evaluated at $z$ = 1.
 Results are shown with respect to the dark matter only non-linear predictions (thick solid line).
 From top to bottom, the (blue) dashed lines represent the effect of massive neutrinos with $M_{\nu}$ = $0.2$, $0.4$ and $0.6$eV respectively.
The combinations of massive neutrinos with baryon feedback are shown with the thin (red) solid lines. }
   \label{fig:Pk_b_nu}
\end{figure}

\begin{figure}
\centering
  \includegraphics[width=2.2in]{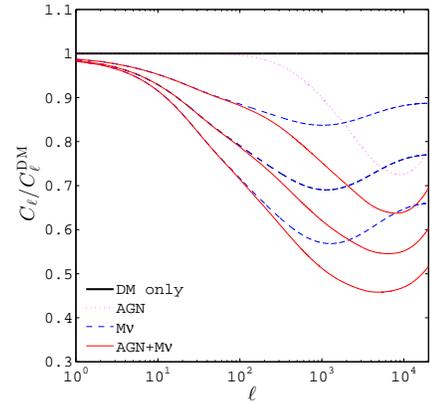} 
\vspace{-0.25cm}
\caption{ Combined effect from baryon feedback and massive neutrinos on the weak lensing power spectrum, assuming the source redshift distribution given by
equation (\ref{eq:nz}) and the baseline {\it WMAP}9 cosmology.
As for Fig. \ref{fig:Pk_b_nu}, results of different combinations are shown with respect to the dark matter only non-linear predictions (thick solid line),
and the sum of neutrino masses shown  are, from top to bottom,  $M_{\nu}$ = $0.2$, $0.4$ and $0.6$eV.}
\label{fig:C_ell_b_nu}
\end{figure}

\begin{figure}
\centering
  \includegraphics[width=3.5in]{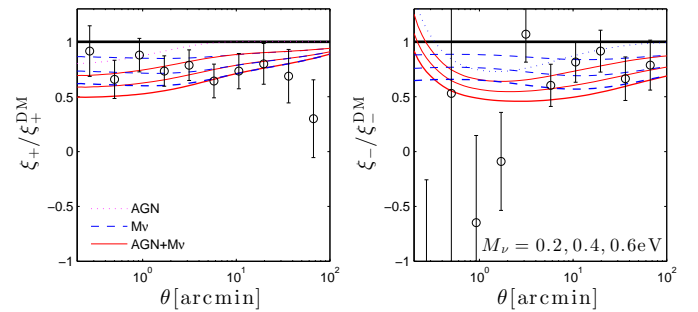} 
\vspace{-0.30cm}
\caption{(left:) Combined effect from baryon feedback  and massive neutrinos on the weak lensing two-point correlation function $\xi_{+}$.
The open symbols represent our measurements from CFHTLenS data, shown with 1$\sigma$ error bars. (right:) Same as the left panel, but for the $\xi_-$ estimator.
We used the same $y$-axis range for both panels to emphasize on the differences across the models, hence the leftmost point falls outside the frame, at $\xi_-/\xi^{\rm DM}_-$=-3.8. 
}
\label{fig:xi_b_nu}
\end{figure}

Fig. \ref{fig:Pk_b_nu} shows the impact of different combinations of baryons and massive neutrinos on the matter power spectrum.
Fig. \ref{fig:C_ell_b_nu} and \ref{fig:xi_b_nu} show the equivalent effects on the weak lensing power spectrum $C_{\ell}^{\kappa}$ and on the shear two-point correlation function $\xi_\pm(\theta)$, respectively.
We can see from the three figures that all models converge to DM-ONLY at large scales (low $k$,  low $\ell$ and high $\theta$), and that the combined effect can suppress more than 50\%
of the power,
depending on the models and neutrino mass. Also, it becomes clear that surveys probing small patches (restricted to $\ell>500$ for example) would have difficulties
to distinguish between the two feedback processes. This degeneracy can only be broken with the inclusion of lower $\ell$ multipoles, where baryon feedback is
minimal but massive neutrinos still leaves a signature \citep{2014arXiv1405.6205N}.

\subsubsection{Combined feedback with modified gravity}
\label{subsubsec:MG_bias}


The evolution of perturbations in the context of large-scale structures has been 
carefully studied in $f(R)$ and scalar-tensors theories
gravity \citep{Zhang06,Koivisto06,Bean06, SHS07, HS07,SPH07,PS08,CDT08,KTH09,MSY09,LH11, Brax11, LZK12,LinMot13,TCP13,BV13}.
In this paper, we focus on the matter density power spectrum $P(k;z)$,
or more precisely, on the weak lensing convergence power spectrum 
$C_{\ell}^{\kappa}$, which can be computed from $P(k;z)$ through the modified Poisson equations
that relate the metric gravitational potentials to the matter density fluctuations.

Therefore, before computing weak lensing statistics, we first need to describe
gravitational clustering and the 3D matter density power spectrum
for all cosmological scenarios that we investigate.
We use the approach first developed in \citet{Valageas2013} for the
$\Lambda$CDM cosmology,  generalized afterwards to various modified-gravity
scenarios in \citet{BV13,BV2014-K2}.
This is an analytical approach that combines perturbation theory up to one-loop
order (i.e., up to order $P_L^2$, where $P_L$ is the linear matter density
power spectrum) with a phenomenological halo model. Namely, 
we are splitting the matter power spectrum as:
\begin{eqnarray}
P(k) = P_{\rm 2H}(k) + P_{\rm 1H}(k) ,
\label{eq:pk}
\end{eqnarray}
where $P_{\rm 2H}(k)$ is the `two-halo' term associated with pairs of particles
that are enclosed in two different halos, whereas $P_{\rm 1H}(k)$ is the
`one-halo' term associated with pairs enclosed in the same halo.
This construction allows us to obtain predictions for the
non-linear matter power spectrum covering the linear, quasi-linear and  highly non-linear scales.
We refer the reader to the work cited above for complete details and validations of equation (\ref{eq:pk}),
but nevertheless provide an overview of the method in the Appendix for quick reference. 
We note that other prescriptions exists for modelling $P(k)$ in modified gravity scenarios, i.e. \citet{2014ApJS..211...23}
for the $f(R)$ model. However the modelling we adopt here applies also to $f(R)$ with $n\ne1$ gravity, to Dilaton gravity, and in fact to any modified gravity model expressed in the tomographic parameterization, 
which makes it general and accurate at the same time.

In analogy with equations (\ref{eq:nu_feedback}) and (\ref{eq:b_feedback}),
we define the {\it modified gravity bias}:
\begin{eqnarray}
b^2_{\rm MG(\alpha)}(k,z) \equiv \frac{P^{\rm MG(\alpha)}_{\rm VNT}(k,z)}{P^{\rm DM}_{\rm VNT}(k,z)},
\label{eq:MG_feedback}
\end{eqnarray}
where $\rm MG(\alpha)$ refers to the gravity model, with $\alpha$= 0 corresponding to GR, $\alpha$= [1, 2, 3, ...,15] specifying dilation models [A1, A2, A3, ...,
E4], $\alpha$= [16, 17, 18] specifying $f(R)$ models with $n=1$ and $|f_{R_0}|$ = $10^{-4}$, $10^{-5}$, $10^{-6}$, and finally $\alpha$= [19, 20, 21] the $f(R)$ models with $n=2$
and the same $|f_{R_0}|$ values.
The subscript `VNT' indicates quantities that are computed in the framework of \citet{Valageas2013}, i.e. with equation (\ref{eq:pk}).

Bringing all the pieces together, we  construct the matter power spectrum for any combination of baryon feedback, neutrino mass and modified gravity by multiplying
the DM-ONLY model by the corresponding biases:
\begin{eqnarray}
  P^{{\rm DM}+\nu+b(\rm m)+{\rm MG}} = P^{\rm DM} \times b^2_{M_{\nu}} \times b^2_{\rm m} \times  b^2_{\rm MG(\alpha)}.
  \label{eq:pk_all}
\end{eqnarray}
We have removed the dependences on scale and redshift for each of these terms to clarify the notation.
This modelling assumes that the effect of modified gravity on the baryon and neutrino feedbacks can be neglected,
allowing for the convenient  factorization presented in equation (\ref{eq:pk_all}).
This seems to be a valid approximation for some models, as it was shown in \citet{Hamani15} that the modified gravity bias measured in dark matter only
matched to better than 5\% the same measurement done in full hydrodynamical simulations, for $f(R)$ models with $n=1$ and $|f_{R_0}| \in [10^{-4} - 10^{-6}]$.
However, the same group also observed larger deviations in many symmetron models, up to 20\% by $k=10h \mbox{Mpc}^{-1}$ in some cases.
This places a limit on the accuracy of equation (\ref{eq:pk_all}), and calls for more hydrodynamical simulation runs where $b_{\rm m}$ and $b_{\rm MG(\alpha)}$
are merged into one term, $b_{\rm m, MG(\alpha)}$, measured for each combination of $\{\alpha, m \}$. This is unfortunately not available at the moment,
hence  equation (\ref{eq:pk_all}) is currently our best shot at this joint measurement. On the neutrino sector, results by \citet{Baldi14} are further encouraging: 
they looked at joint simulations of modified gravity and massive neutrinos and came to the conclusion that one could consider the effect of each almost independently,
supporting the validity of equation (\ref{eq:pk_all}).

\begin{figure}
\centering
  \includegraphics[width=3.35in]{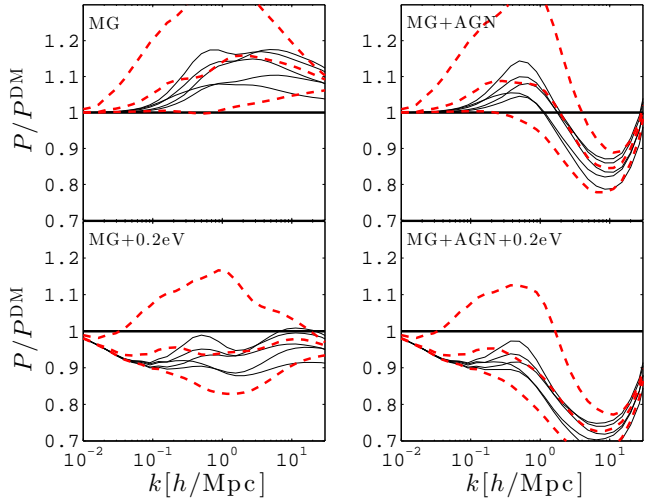} 
\vspace{-0.25cm}
\caption{Combined effect from baryon feedback and massive neutrinos on the matter power spectrum $P(k)$ assuming different modified gravity models, again
evaluated at $z = 1$.
 Results are shown with respect to the dark matter only non-linear predictions (thick horizontal solid line).
 From top to bottom at $k=0.2h\mbox{Mpc}$, the solid lines represent Dilaton models B4, A3, E3, D1 and C1 respectively.
 The thick red dashed lines correspond to $f(R)$ gravity with $n=1$. 
 Top to bottom are for $|f_{R_0}|$ = $10^{-4}$, $10^{-5}$ and $10^{-6}$ respectively.
 We do not show the $n=2$ results to avoid over-crowding the figure, 
 but they are qualitatively similar in shape to the $n=1$ case, albeit with a smaller departure from $\Lambda$CDM.
Different panels show different combinations of massive neutrinos and baryon feedback on these same models, all computed with equation (\ref{eq:pk_all}). }
\label{fig:Pk_b_nu_mg}
\end{figure}

\begin{figure}
\centering
  \includegraphics[width=3.35in]{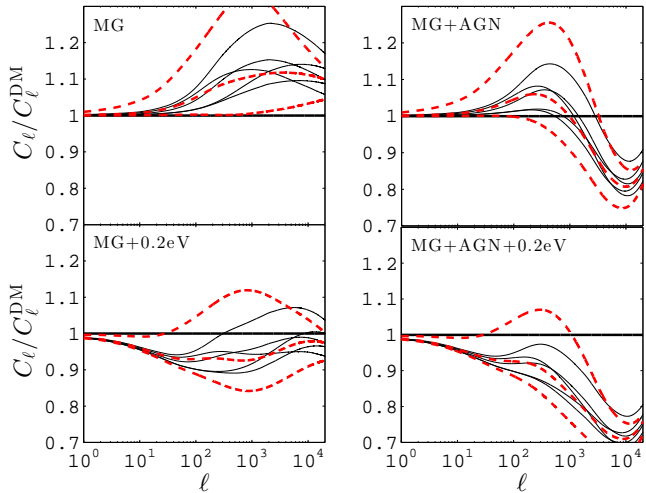} 
\vspace{-0.25cm}
\caption{Same as Fig. \ref{fig:Pk_b_nu_mg}, but for the weak lensing power spectrum.}
\label{fig:Cell_b_nu_mg}
\end{figure}

For each combination, we compute predictions for the weak lensing quantity with equations (\ref{eq:limber}) and (\ref{eq:xi}).
We report our results on $P(k)$ and $C_{\ell}^{\kappa}$ in Fig. \ref{fig:Pk_b_nu_mg} and \ref{fig:Cell_b_nu_mg} respectively.
Whereas modified gravity is generally boosting the clustering compared to a $\Lambda$CDM universe, the inclusion
of massive neutrinos and/or baryonic feedback  is working in the opposite direction.
It becomes clear that a precise distinction between these three feedback contributions  poses a challenge to clustering and weak lensing experiments.

\subsection{Data}

Our measurement of the shear correlation functions $\xi_{\pm}$ is based on the public release of the Canada France Hawaii Telescope Lensing Survey
(CFHTLenS\footnote{CFHTLenS: {\tt www.cfhtlens.org}}). The CFHTLenS covers a total area of 154 deg$^2$, which is reduced to
128 deg$^2$ after masking bright stars, foreground moving objects and faulty CCD rows.
Full details about the data reduction pipeline are provided in \citet{2013MNRAS.433.2545E}.
Source redshifts are obtained from the five bands $u'griz$  photometric observations \citep{2012MNRAS.421.2355H} and were 
carefully tested in \citet{2013MNRAS.431.1547B};
shape measurements are performed on the $r$-band images  with the {\it lens}fit Bayesian code  described in \citet{2013MNRAS.429.2858M}.
A detailed assessment of the residual systematics is provided in \citet{2012MNRAS.427..146H}, and we refer the reader to these
references for more information on the CFHTLenS data.

As described in \citet{2012MNRAS.427..146H}, the public shear data must be recalibrated with additive and multiplicative factors,
commonly referred to as the $c$ and $m$  corrections. In contrast with this reference, we use a different $c$ correction, as detailed in HWVH,
which is less model dependent. Although the overall change on the correction is marginal, the number of CFHTLenS pointings that are flagged as {\it bad}
is reduced by almost a half.

{

Following the recommendations of  \citet{2012MNRAS.427..146H} and \citet{2013MNRAS.431.1547B}, we  minimize the systematic contamination
from  badly reconstructed photometric redshifts by applying the selection cut  $0.4<z_{\rm phot}<1.3$. 
We construct the redshift distribution $n(z)$ for the selected galaxies from the {\it lens}fit-weighted stacked
probability distribution functions of the galaxy sample. As shown in HWMH, the distribution is well described  by the following analytical expression:
%
%
%
\begin{eqnarray}
n(z)=N_0 e^{-(z-z_0)^2/\sigma_0^2}+ N_1 e^{-(z-z_1)^2/\sigma_1^2} \nonumber \\
+  \frac{N_2 e^{-(z-z_2)^2/\sigma_2^2}}{1.0+e^{-10.0 (z-0.6)}},
\label{eq:nz}
\end{eqnarray}
where $(N_0,z_0,\sigma_0,N_1,z_1,\sigma_1,N_2,z_2,\sigma_2)$ = (
0.14438,
     0.760574,
     0.14594,
     0.514894,
     0.498379,
     0.15608,
     1.74435,
     0.445019,
     0.684098).
There is a $0.4$ percent difference in the mean redshift between the fit and the distribution, which yields a small error well below the
other sources of error in our analysis. We therefore neglected this contribution to the systematic budget.


We construct our shear correlation function estimator following \citet{2013MNRAS.430.2200K}:
\begin{eqnarray}
\xi_{\pm}(\theta)={\sum_{i,j} w_i w_j \left[e_t(\theta_i) e_t(\theta_j) \pm e_r(\theta_i) e_r(\theta_j)\right]\over \sum_{i,j} w_i w_j}.
\label{eq:xi_estimator}
\end{eqnarray}
All galaxy pairs $(i,j)$ separated with angular distance $|\theta_i-\theta_j| \in \theta$ contribute to the same bin,
with their contribution weighted by the product of their {\it lens}fit weights $w_i w_j$ \citep{2013MNRAS.429.2858M}.
The shear quantities $e_t$ and $e_r$ are the tangential and cross-component of the galaxy ellipticity,
measured in the coordinate system of the galaxy pair.
We account for the shear calibration by measuring
\begin{eqnarray}
1+K(\theta)={\sum_{i,j} w_i w_j (1+m_i)(1+m_j)\over \sum_{i,j} w_i w_j}
\end{eqnarray}
and dividing  $\xi_{\pm}$ by $1+K$. As a rule of thumb, $K$ is $\sim-0.11$ at all angular scales, with variations smaller than 0.1\%.
We finally exclude all pairs with $\theta<12$ arcseconds in order to
minimize contamination by post stamp leakage across {\it lens}fit templates.
We perform this measurement with {\small ATHENA}\footnote{ATHENA: {\tt http://cosmostat.org/athena.html}},
and show our results in Fig. \ref{fig:xi_b_nu}.

\subsection{Simulations}

In order to achieve a high precision cosmic shear measurement, not only must the data be thoroughly tested for subtle systematics residuals,
but the sampling variance must be accurately estimated, a quantity that is very hard to assess from the data.
To overcome this difficulty, we rely on a suite of weak lensing simulations based on {\it WMAP}9 + SN + BAO cosmology.
As detailed in \citet{2014arXiv1406.0543H}, the SLICS-LE suite consists of $60$ sq. degrees light cones extracted from 500 independent $N$-body
realizations.
The numerical weak lensing signal is precise to better than 10 percent for $\xi_+$ with $\theta > 0.4'$ (and $\theta > 5'$ for $\xi_-$).
We construct the mock maps by combining the different redshift planes with a redshift source distribution that mimics that of the data.
We then sample the simulated shear maps with $10^{5}$ points randomly located, and compute the shear two-point correlation functions $\xi_{\pm}$
of these mock `galaxies' with the same pipeline as the data (i.e. from equation \ref{eq:xi_estimator}).

\begin{figure}
\centering
  \includegraphics[width=3.35in]{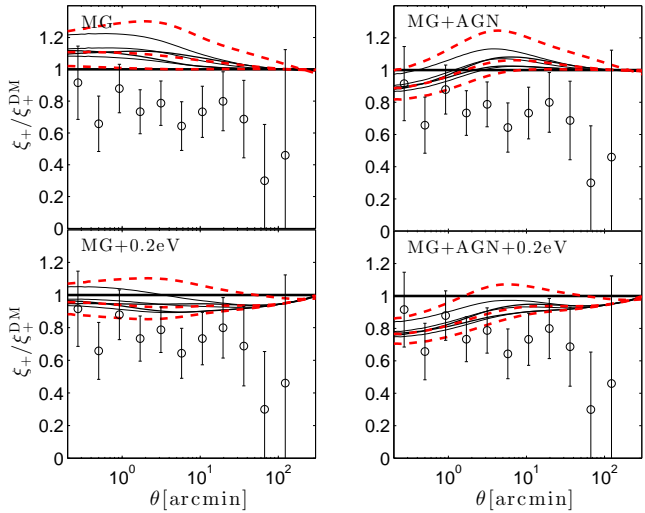} 
\vspace{-0.25cm}
\caption{Same as Fig. \ref{fig:Pk_b_nu_mg}, but for $\xi_+$. The open symbols represent our measurements from the CFHTLenS data, exactly as in the right panel of Fig. \ref{fig:xi_b_nu}.
Shown are the 1$\sigma$ error bars.}
\label{fig:xip_b_nu_mg}
\end{figure}

\begin{figure}
\centering
  \includegraphics[width=3.35in]{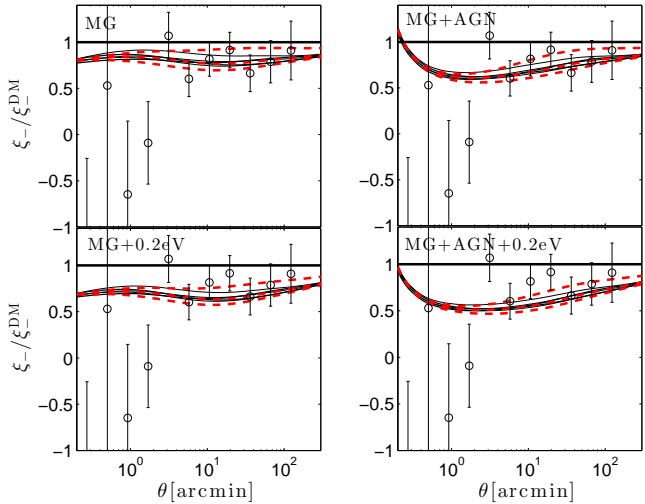} 
\vspace{-0.25cm}
\caption{Same as Fig. \ref{fig:xip_b_nu_mg}, but for $\xi_-$. Note the different  $y$-axis range compared to Fig. \ref{fig:xip_b_nu_mg}.}
\label{fig:xim_b_nu_mg}
\end{figure}

\section{Results}
\label{sec:results}

In this Section,  we first review our error budget, we then describe how different components combine in our model rejection procedure,
and finally present our results.

\subsection{Error budget}
\label{subsec:error}

This analysis closely  follows that of HWVH;  we summarize here the main ingredients, and refer the reader to the reference for more details.
The sources of error in this analysis can be broken into three terms:
1- uncertainty on the cosmic shear measurement, 2- uncertainty in the theoretical model describing the non-linear regime of structure formation, and 3- uncertainty
on the fiducial cosmology.

1- The error on our cosmic shear measurement is dominated by shape noise at small angles and sampling variance at large angles.
The angular scales at which these two errors contribute equally occur at $\theta=2$ and $30$ arc minutes for $\xi_+$ and $\xi_-$, respectively.
In addition, the variance-shape noise mixed term contributes to about a third of the error on $\xi_+$ at large angles, but is negligible in $\xi_-$, as seen in
\citet{2013MNRAS.430.2200K}.
We have estimated the sampling variance from the SLICS-LE weak lensing simulations, and added an extra contribution from the {\it halo sampling variance}, following
the modelling of
\citet{2009ApJ...701...945S}, which provides at most a 10 percent correction on the overall error.
Our measurement is minimally affected by intrinsic alignment of galaxies, since we do not perform tomographic analysis \citep[see][for more details on intrinsic alignments in the CFHTLenS data]{2013MNRAS.432.2433H}.
The error from shape reconstruction is already included in the statistical term, hence does not require an extra term.
Photometric redshift uncertainty enters the measurement through modification of the source distribution $n(z)$, but
this effect is negligible compared with other sources of error hence is not included.

2- The uncertainty on the dark matter only non-linear model has been carefully assessed in HWVH by comparing five different predictions:
{\small HALOFIT}2012, {\small HALOFIT}2011 + small scale empirical recalibration, Cosmic Emulator + power law graft, Cosmic Emulator + {\small HALOFIT}2012 graft
and, finally, the mean over five independent high resolution simulations -- the SLICS-HR suite described in  \citet{2014arXiv1406.0543H}.
These five models agree very well over most angular scales, and the 1$\sigma$ scatter among them is taken as the ($\theta$-dependent) theoretical error.

3- The uncertainty in the cosmological parameters is set by the {\it WMAP}9 precision \citep{2013ApJS..208...19H}, whose dominant contribution on the weak lensing uncertainty
arises via the parameters $\Omega_M$ and $A_s$.
With the inclusion of the BAO and SN external data, these two parameters are allowed a  3.4 and 3.3 percent variation about their mean values (1$\sigma$).
Since the amplitude of $\xi_{\pm}$ roughly scales as $(A_s \Omega_M)^2$, we expect the combined error to be of the order 5 percent of the $\Lambda$CDM baseline signal, assuming no prior on the joint
contour.

Note that the cosmological error and the modelling error (terms 2 and 3) enter in our analysis  as systematic uncertainties, therefore we add them in quadrature and marginalize over them (see details in \textsection\ref{subsec:model_rejection}). Also note that the Planck cosmology $\{\Omega_M, A_s \}$ falls within our $3\sigma$ search limits, although closer to the edge of the search zone.

\subsection{Model rejection strategy}
\label{subsec:model_rejection}

As seen in Fig. \ref{fig:xip_b_nu_mg} and \ref{fig:xim_b_nu_mg}, the effects of baryons, massive neutrinos and modified gravity
 are significantly degenerate on $\xi_{\pm}$. Given the noise levels in the current data  and the number of internal parameters that describe these different mechanisms, 
 performing a full MCMC analysis is not convenient to extract meaningful constraints.
A more appropriate and direct way is to sample a finite set of model combinations and examine their agreement with the data.
This case-by-case strategy has the potential to reject models that are inconsistent with the data,
which can then be translated into constraints on the underlying free parameters.

The metric we adopt for this type of analysis is the $p$-value, which
measures the probability that the data is consistent with the model, {\it if the model is true}.
It is given by the integral over the $\chi^2$ probability density function, where the lower bound is the measured $\chi^2$ and the upper bound is infinity.
As a rule of thumb, models with $p$-values $<10\%$ are rejected with more than 90\% confidence,
and $1\sigma, 2\sigma, 3\sigma...$ rejection measurements are obtained for $p$-values of 0.317, 0.046, 0.003....
Our strategy therefore consists to measure the $\chi^2$ and $p$-value associated with each combination of baryon feedback, neutrino mass and
modified gravity model, and to flag every combination with $p<0.1$ as being disfavoured.

The uncertainty arising from statistical and sampling variance naturally enters this calculation through the evaluation of the $\chi^2$,
which involves the inversion of the cosmic shear covariance matrix.
The systematic uncertainty, however, is trickier to capture.
In our cosmic shear measurement, it mainly manifests itself as shifts in the amplitude of the signal, as described in \textsection\ref{subsec:error}.
The systematic error  is higher at smaller angles and represents at most an error of $\sim9\%$ on the $\xi_{\pm}$ model amplitude.
In order to marginalize over this effect, for each model, we allow the theoretical signal to shift up and down by $3\sigma_{\rm syst}$,
corresponding to vertical excursion of $27\%$ in Fig. \ref{fig:xip_b_nu_mg} and \ref{fig:xim_b_nu_mg},
keeping the error bars (statistical + sampling) fixed.
We then compute an array of $p$-values in this excursion range, and record only the largest measurement (i.e. the least restrictive).

The exact number of degrees of freedom ({\it d.o.f.}) that enters the $\chi^2$ distribution function must be carefully chosen.
To begin with, each of the two cosmic shear observables  is organized in 11 angular bins,
yielding a maximum of 22 {\it d.o.f.}
However, assigning one {\it d.o.f.} per angular bin would be incorrect, for the following reason.
In a statistical sense, our model rejection method is completely equivalent to fitting the parameter combination $(A_s^2\Omega_M^{1.8})$ from the amplitude of the
$\xi_{\pm}$ signals,
followed by an extraction of the most likely neutrino mass for each baryon feedback and modified gravity model from the largest $p$-value.
This implies that the number of degrees of freedom should be reduced by two (one for fitting $A_s^2\Omega_M^{1.8}$, one for fitting $M_{\nu}$) in the conversion
between $\chi^2$ and $p$-values.

Note that for a given angular scale, both $\xi_+$ and $\xi_-$  probe different physical scales, the latter focusing on structures about five times smaller.
It is therefore relevant to examine the constraining power of $\xi_+$ first, and to add $\xi_-$ to the data vector as a second step.
When both are combined, the full data covariance matrix involves the cross-correlation region, as described in HWVH.
The resulting $p$-values are summarized for all our results in Table \ref{table:pvalues}, for $M_{\nu} \le 0.2$ eV.
No conclusions can be drawn from models with higher total neutrino masses, as the $p$-values for any combination is always  greater than 0.175.
The models rejected at more than $1.64\sigma$ (i.e. 90\% CI) are highlighted in bold font.

\begin{table*}
\centering
\caption{Distribution of $p$-values for different combination of baryon feedback models, neutrino masses and gravity models (see main text for details). 
 The parameters listed in the leftmost column of the $f(R)$ models are $\{n, |f_{R_0}| \}$. Dilaton models  are described in  Table \ref{tabular:tab2}.
For this calculation, we
fit all data in the range $0.2 < \theta < 167$ arcminutes.
Specifically, each entry in this Table represents the largest $p$-value probed inside a $3\sigma_{\rm syst}$ region about the mean of the model.
Values in bold face highlight the model combinations that are excluded by the data with more than $1.64 \sigma$ significance ($p$-value $<0.1$, equivalent to a
confidence interval (CI) of 90\%). Models with $M_{\nu} > 0.2$ eV are not listed, as none has value lower than 0.176.}
\begin{tabular}{@{} |c|cc|cc|cc|cc| @{}c} 
\hline
\hline
& \multicolumn{2}{c|}{DM-ONLY} &  \multicolumn{2}{c|}{AGN}&  \multicolumn{2}{c|}{0.2eV}&  \multicolumn{2}{c|}{AGN+0.2eV}\\
Model& $\xi_+$ & $\xi_+\xi_-$ & $\xi_+$ & $\xi_+\xi_-$ & $\xi_+$ & $\xi_+\xi_-$ & $\xi_+$ & $\xi_+\xi_-$ \\
\hline
\hline
 \multicolumn{9}{|c|}{General Relativity}\\
 \hline
      $\Lambda$CDM  & 0.132 & \bf0.065 & 0.119 & 0.150 & 0.331 & 0.297 & 0.289 & 0.444 \\
\hline
\hline
 \multicolumn{9}{|c|}{Generalized Dilaton}\\
 \hline
  A1  & 0.126 & \bf0.058 & 0.116 & 0.141 & 0.323 & 0.282 & 0.284 & 0.431 \\
  A2  & \bf0.088 & \bf0.030 & \bf0.093 & \bf0.099 & 0.269 & 0.203 & 0.256 & 0.363 \\
  A3  & \bf 0.037 & \bf0.008 & \bf0.060 & \bf0.051 & 0.184 & 0.105 & 0.207 & 0.265 \\
  B1  & 0.120 & \bf0.054 & 0.113 & 0.135 & 0.315 & 0.273 & 0.281 & 0.424 \\
  B3  & \bf 0.043 & \bf0.010 & \bf0.064 & \bf0.054 & 0.195 & 0.113 & 0.212 & 0.272 \\
  B4  & \bf 0.008 & \bf0.001 & \bf0.032 & \bf0.019 & 0.107 & \bf0.037 & 0.150 & 0.161 \\
  C1  & \bf 0.080 & \bf0.022 & 0.100 & \bf0.098 & 0.259 & 0.171 & 0.271 & 0.365 \\
  C3  & 0.104 & \bf0.040 & \bf0.098 & 0.111 & 0.293 & 0.239 & 0.261 & 0.385 \\
  C4  & 0.114 & \bf0.049 & 0.104 & 0.122 & 0.307 & 0.259 & 0.267 & 0.402 \\
  D1  & \bf 0.063 & \bf0.013 & 0.100 & \bf0.085 & 0.232 & 0.132 & 0.276 & 0.347 \\
  D3  & 0.127 & \bf0.060 & 0.115 & 0.141 & 0.325 & 0.286 & 0.282 & 0.431 \\
  D4  & 0.131 & \bf0.064 & 0.119 & 0.149 & 0.331 & 0.295 & 0.288 & 0.441 \\
  E1  & 0.118 & \bf0.049 & 0.117 & 0.135 & 0.314 & 0.260 & 0.289 & 0.425 \\
  E3  & \bf0.047 & \bf0.013 & \bf0.053 & \bf0.049 & 0.200 & 0.132 & 0.186 & 0.257 \\
  E4  & \bf0.026 & \bf0.007 & \bf0.032 & \bf0.027 & 0.156 & \bf0.094 & 0.138 & 0.188 \\
  \hline
  \hline
 \multicolumn{9}{|c|}{$f(R)$}\\
 \hline
  $\{ 1,10^{-4} \}$  & \bf0.001 & \bf0.000 & \bf0.005 & \bf0.003 & \bf0.051 & \bf0.018 & \bf0.054 & \bf0.057 \\
  $\{ 1,10^{-5} \}$  & \bf0.058 & \bf0.013 & \bf0.072 & \bf0.062 & 0.222 & 0.134 & 0.222 & 0.292 \\
  $\{ 1,10^{-6} \}$  & 0.129 & \bf0.054 & 0.125 & 0.145 & 0.328 & 0.271 & 0.298 & 0.437 \\
  $\{ 2,10^{-4} \}$  & \bf0.011 & \bf0.003 & \bf0.020 & \bf0.014 & 0.112 & \bf0.056 & 0.104 & 0.131 \\ 
  $\{ 2,10^{-5} \}$  & \bf0.095 & \bf0.030 & \bf0.094 & \bf0.097 & 0.277 & 0.200 & 0.254 & 0.354 \\ 
  $\{ 2,10^{-6} \}$  & 0.137 & \bf0.063 & 0.126 & 0.154 & 0.338 & 0.292 & 0.299 & 0.449 \\ 

\hline
\hline
   \end{tabular}
  \label{table:pvalues}
\end{table*}

\subsection{Discussion}
\label{subsec:discussion}

One of the main results recovered from Table  \ref{table:pvalues} is that the $f(R)$ model with $\{|f_{R_0}|,n\}= \{10^{-4}, 1\}$ is strongly disfavoured by the cosmic shear data,
regardless of the baryonic feedback model or sum of neutrino mass, consistent with independent constraints.
The $f(R)$ and $f(R)$ + AGN models are rejected by at least $3\sigma$, but combinations including massive  (0.2eV) neutrinos tend to weaken this constraints.
This can be understood by the fact that massive neutrinos and modified gravity partly compensate for one another, reducing the global departure from $\Lambda$CDM.
Also, $f(R,n$=$2)$ models are generally in better agreement with the data compared to their $(n$=$1)$ counterpart. 
This is so simply because  higher values of $n$ rapidly suppress the $f(R)$ term, hence deviations from GR, as seen in equation (\ref{eq:fofR}).  

The next important result is that the rejection of massless neutrinos + DM-ONLY is robust against all modified gravity models we have  tested,
and typically made stronger. The cosmic shear data clearly prefers lower values of $\xi_{\pm}$ at small angular scales, and modified gravity pulls the other way.

The inclusion of baryon feedback reduces to about two thirds the number of models rejected with 90\% CI.
For instance, dilation models A2, A3, B3, B4, C1, C3, D1, E3 and E4 are disfavoured; these are the most discrepant with GR+$\Lambda$CDM. 
Referring to Table \ref{tabular:tab2} and the model descriptions in \textsection\ref{sec:Dilatons},  this can be interpreted as follow.
In a tomographic parameterization of modified gravity centred on $\{m_0, r,  \beta_0, s\} = \{0.334, 1.0, 0.5,  0.24 \}$, excursion
in the $s$, $\beta_0$  and $r$ directions are studied with models A, B and C respectively, 
and the data favours lower parameters values. Model E explores the $m_0$ direction, and the data prefers higher values.
Model D explores the diagonal direction in the $\{m_0, s\}$ plane at fixed $A_2$ (see equation \ref{eq:s-def}), where here we observe instead that the data prefers lower $m_0$ values.

We note that there is a mild effect seen in the `AGN' column of Table \ref{table:pvalues}, where the addition of $\xi_-$ to the data vector sometimes increases the $p$-value by a small amount.
This can be attributed to the fact that at small angles, $\xi_-$ prefers amplitude even lower than $\xi_+$, compared to the DM-ONLY model.
Adding baryon feedback therefore produces a lower rejection rate in the former than in the latter quantity.

When neutrino masses are allowed to reach 0.2eV, only the $f(R) \{10^{-4},1\}$, $f(R) \{10^{-4},2\}$ and the Dilaton B4 and E4 models remain in tension with the data.
With AGN + $M_{\nu}$=0.2eV, no models are rejected, aside from  the most extreme case considered in this paper: $f(R) \{10^{-4},1\}$.

This means that given the current cosmic shear data and levels of systematics, it is possible to accommodate most models, as long as 
either massive neutrinos or strong baryon feedback mechanisms counter-balance the effect of the fifth force on the matter clustering.
As upcoming independent cosmological probes will tighten the uncertainty on neutrino masses and significantly improve the statistical and sampling errors,
we expect the next generation of such analysis to be much more constraining. 
Once at this stage, it will be instructive to propagate our measurements 
onto $\{m(a), \beta(a) \}$ contours and provide a Fisher  matrix for joint probes analyses.
If, for instance, the total mass turns out to be much smaller than 0.2eV,
then the current  AGN column should gives a very good approximation of the rejection power from the CFHTLenS cosmic shear data.
Precise modelling of the baryon feedback is likely to take more time to reach, due to the higher level of complexity intrinsic to these astrophysical phenomena. 
Intermediate solutions will involve a series of tuneable parameters, also to be constrained.

On a separate note, we stress that the constraints can be further tightened using additional information about the weak-lensing observables, such as the non-Gaussian features \citep{2012MNRAS.419..536M},
or by combining the results with external probes such as redshift distortions, peculiar velocity, etc.

\section{Conclusion}
\label{sec:conclu}

Cosmic shear is a promising tool for probing deviations from GR, since these are maximal at scales of a few Mpc,
where the lensing signal-to-noise ratio is the highest.
These same scales are very challenging to probe with other types of large scale structure observables, mainly because of the galaxy bias that is largely unknown.
At the same time, this complimentarity offers a number of opportunities for strong constraints based on joint data sets. 

One of the main challenge in working with these non-linear scales is the large theoretical uncertainties due to the unknown neutrino masses, the precise baryonic feedback mechanisms 
and, to a lesser extend, inaccuracies  in the clustering of dark matter. However, a lot of effort is invested in all these areas, such that it becomes possible to place 
joint constraints on these degenerate physical effects.

This paper presents the first constraints on modified gravity obtained from cosmic shear measurements alone; 
the results are derived by studying the impact of modified gravity on matter clustering and comparing the predictions with the public CFHTLenS data.
Limiting the background $\Lambda$CDM cosmology to the $3\sigma$ range in $\{A_s, \Omega_M\}$ allowed by {\it WMAP}9 + SN + BAO,
we compared the $\xi_{\pm}$ data against predictions  including $f(R)$ and Dilaton models in a number of parameter configurations.
We carried a model rejection analysis  accounting for possible degeneracies with massive neutrinos and baryonic feedback mechanisms,
and investigated which combinations of models were mostly disfavoured by the data.
 As summarized in Table \ref{table:pvalues},
the $f(R)$ model with $|f_{R_0}|=10^{-4}$ is strongly disfavoured even in the presence of realistic levels of  baryonic  feedback and massive neutrinos reaching $M_{\nu}$=0.2eV.
A universe with no baryonic feedback and massless neutrinos is also rejected with $2\sigma$ or above in most modified gravity scenarios.
We are not yet able to identify a preferred model with the current level of statistical accuracy, but we expect future weak lensing experiments
to improve significantly in this direction.

In our analyses, we have use the simplifying assumption that the biases due to massive  neutrinos, baryon feedback and modified gravity were uncorrelated, 
which is justified to some extend based on the several numerical results. However, precise correlations will need to be studied for a number 
of models, a task that involves suites of large cosmological hydrodynamical simulations including all the ingredients at once.

One important future tasks will be to map observational detections of modifications to GR onto parameter constraints such as the $\{m(a), \beta(a) \}$ pair.
However, the current data is not quite there yet.
Several theories can accommodate similar phenomenological effects, and model-independent parameterizations such as that presented in \citet{LBF15} 
  might prove helpful for this.

This paper used the impact of modified gravity on the clustering properties of matter and their propagation onto the weak lensing cosmic shear signal.
Other avenues of probing deviations from GR with weak lensing data are complimentary, including direct combinations with baryonic probes
or tomographic decomposition, and worth exploring in the near future.

\section*{Acknowledgements}
We would like to thank Alexander Mead for his comments on the manuscript, 
and to acknowledge useful discussions with Alexei Starobinsky, Alan Heavens, Catherine Heymans and Massimo Viola.
We are deeply grateful to all the CFHTLenS team for having made public their high quality shear data.
Computations for the $N$-body simulations were performed on the GPC supercomputer at the SciNet HPC Consortium.
SciNet is funded by: the Canada Foundation for Innovation under the auspices of Compute Canada;
the Government of Ontario; Ontario Research Fund - Research Excellence; and the University of Toronto.
This work is based on observations obtained with MegaPrime/MegaCam, a joint project of CFHT and CEA/IRFU,
at the Canada-France-Hawaii Telescope (CFHT) which is operated by the National Research Council (NRC) of Canada,
the Institut National des Sciences de l'Univers of the Centre National de la Recherche Scientifique (CNRS) of France,
and the University of Hawaii. This research used the facilities of the Canadian Astronomy Data Centre operated
 by the National Research Council of Canada with the support of the Canadian Space Agency.
 CFHTLenS data processing was made possible thanks to significant computing support from the NSERC Research Tools and Instruments grant program.
JHD and LvW are funded by the NSERC and Canadian Institute for Advanced Research CIfAR;
DM and PC acknowledge support from the Science and Technology Facilities Council (grant numbers ST/L000652/1; 
LR and PV receive support from the French Agence Nationale de la Recherche
under Grant ANR-12-BS05-0002;
PB acknowledges partial support from the European Union FP7 ITN
INVISIBLES (Marie Curie Actions, PITN- GA-2011- 289442) and from the Agence Nationale de la Recherche under contract ANR 2010
BLANC 0413 01.


\bibliography{paper.bbl}

\appendix

\section*{}

\subsection*{APPENDIX: Details on the Theoretical Modelling of $P(k)$ in Modified Gravity Scenarios}
\label{app:1}

This Appendix discusses the construction strategy of the matter density power spectrum  $P(k;z)$ in the presence of $f(R)$ or Dilaton modifications 
to General Relativity; full details are provided in the references contained herein.

\subsection{Two-halo term: $P_{\rm 2H}(k)$}
\label{sec:two-halo}

The power spectrum analytical prediction is constructed from a halo model approach, following equation (\ref{eq:pk}).
The two-halo term, which dominates on large scales, is computed
from a Lagrangian-space resummation of standard perturbation theory
that is exact up to order $P_{\rm L}^2$ and contains partial resummations of
higher order terms.
It is also supplemented with non-perturbative contributions
that take into account some aspects of shell crossing 
and ensure that all particle pairs are counted only once in the sum.
Within this framework, the large-scale term $P_{\rm 2H}(k)$ essentially
contains no free parameter.
It can therefore be computed in $\Lambda$CDM and modified-gravity
scenarios by using perturbation theory up to order $P_{\rm L}^2$ (which requires going to order $\delta_L^3$ in terms of the density field
itself).

In the case of the $\Lambda$CDM cosmology, this perturbative expansion
follows the standard approach \citep{BCGS2002}, where the density 
and velocity fields
are written as perturbative expansions over powers of the linear density
field $\delta_L$;  subsequent orders are computed by substituting into 
the continuity and Euler-Poisson equations.
In the case of the modified-gravity scenarios  considered in this paper,
we follow the same approach but require an additional
expansion to write the fifth force in terms of the non-linear density fluctuations.
Indeed, using the quasi-static approximation,  we can relate the scalar field $\varphi$
to the matter density field $\rho$, typically through a non-linear
Klein-Gordon equation. Then, we can solve for $\varphi$ 
as an expansion over the non-linear density fluctuations $\delta\rho$.
This allows us to obtain both the Newtonian potential and the
fifth-force potential 
as functionals of the non-linear matter density fluctuations.
However, while the Newtonian potential is given by the linear Poisson
equation, the fifth-force potential is usually given by a non-linear
equation that involves new time and scale dependences.
In terms of the diagrammatic expansion of the non-linear power spectrum
$P(k)$ over $P_L(k)$, this implies that the linear propagators and the vertices 
are modified with 
new diagrams associated with the new non-linearity of the modified
Poisson equation. See \citet{BV13} for more explanations.

\subsection{One-halo term: $P_{\rm 1H}(k)$}
\label{sec:One-halo}

The one-halo term is obtained from the halo mass function and
the halo density profile, with the addition of a counter-term first introduced
in \citet{VN2011} that arises from mass conservation.
This also ensures that $P_{\rm 1H}(k)$ decays at low $k$ and becomes
subdominant as compared with $P_{\rm 2H}(k)$, whereas the
usual formulation gives a spurious white-noise tail that dominates
on very large scales. 
We take into account the impact of modified gravity through its effect on 
the halo mass function (i.e., through the acceleration or slowing down of the 
spherical collapse), but neglect the impact on the halo shape and profile. 
This should be sufficient for our purposes, because we only consider 
cosmologies that remain close to the $\Lambda$CDM reference, and
these modified gravity models have a much stronger impact on the
halo mass function, especially on its large-mass tail, than on the halo
profile. As shown in \citet{V2013-acc},
at $z=0$, a $10\%$ change to the mass-concentration relation only yields
a $2\%$ change of $P(k)$ at $1 h$Mpc$^{-1}$, 
whereas a $10\%$ change to the halo mass function yields a 
$2\%$ change of $P(k)$ at $0.35 h$Mpc$^{-1}$ and 
a $7.5\%$ change at $1 h$Mpc$^{-1}$.
Generally, the concentration parameter always remains in the
range $3-10$ for typical halos and does not vary by much more than
$10\%$ for realistic scenarios, whereas the mass function
at $M \sim 5\times 10^{14}h^{-1} M_{\odot}$ can vary by more than $50\%$ \citep{LKZL2012, LLKZ2013}.
The interior of haloes are mostly affected by screening anyway, further justifying this approximation.

\subsection{Comparison with numerical simulations}
\label{sec:comparison-Pk}

The modelling described above for the matter density power spectrum 
has been checked in
details against numerical simulations in \citet{Valageas2013} for $\Lambda$CDM 
cosmologies, and in \citet{BV13} for the class of  modified gravity models that we consider
in this paper.
In the case of $\Lambda$CDM, it provides
an accuracy of $2\%$ up to comoving wavenumber $k \sim 0.9 h \rm{Mpc}^{-1}$, 
and $15\%$ up to $k=15 h \rm{Mpc}^{-1}$ down to $z=0.35$.
 In terms of the real-space correlation 
function, this translates into an accuracy of $5\%$ down to the comoving scale $r=0.15 h^{-1} \rm{Mpc}$.
For the $f(R)$ theories, it is able to reproduce very well the deviations from the
$\Lambda$CDM scenarios up to $k=3 h \rm{Mpc}^{-1}$ (the highest wave number
available from the simulations) at $z=0$, for $|f_{R_0}|=10^{-4}, 10^{-5}$, and
$10^{-6}$. In particular, it accurately captures the relative effect on the power compared to the 
$\Lambda$CDM reference due to the non-linear Chameleon mechanism.
For the Dilaton models, the agreement with the numerical simulations
depends somewhat on the model parameters but it typically gives a good quantitative
estimate of the deviations from $\Lambda$CDM up to $k=5 h \rm{Mpc}^{-1}$
(the highest wave number available from the simulations).
When there is a noticeable departure from the simulations, it corresponds to an
underestimation of the amplification of the power spectrum at $k \gtrsim 2 h \rm{Mpc}^{-1}$,
which may be due to our neglect of the impact of modified gravity on the halo concentration
parameter.
Therefore, in such cases our approach provides a conservative estimate of the
deviations from $\Lambda$CDM.
Again, this modelling is able to capture the decrease of the deviations from the
$\Lambda$CDM reference due to the non-linear Damour-Polyakov mechanism.

\end{document}